\documentstyle[aps,preprint,tighten]{revtex}

\begin{document}
\draft

\title{$g_{\pi \Lambda \Sigma}$ and $g_{K\Sigma \Xi}$ from QCD sum rules}
\author{Seungho Choe\thanks{E-mail: schoe@physics.adelaide.edu.au}}
\address{Special Research Centre for the Subatomic Structure of Matter\\
University of Adelaide, Adelaide, SA 5005, Australia.}
\maketitle
\begin{abstract}

The coupling constants
$g_{\pi \Lambda \Sigma}$ and
 $g_{K\Sigma \Xi}$ are calculated in the QCD sum rule approach
 using the three-point function method and 
taking into account  the SU(3) symmetry breaking effects.
The pattern of SU(3) breaking
appears to be different from that based on SU(3) relations.

\end{abstract}
\vspace{1cm}

\section{Introduction}

Understanding the explicit SU(3) symmetry breaking effects in physical 
quantities, such as mass splitting, coupling constants and decay constants, 
has been a subject of research in models of QCD for many years.   
Among those models, the method of
QCD sum rules\cite{SVZ79,RRY85,Narison89}
 has proved to be
a very effective tool to extract information about hadron properties.
In the QCD sum rule approach
the SU(3) breaking effects are
included systematically in perturbative quark mass corrections
(i.e., $m_u = m_d \neq m_s$)
and the different quark condensates ($\langle\bar{u}u\rangle =
\langle\bar{d}d\rangle \neq \langle\bar{s}s\rangle$).
 From fitting analyses of meson and baryon mass splittings
it was found that 
the best fit was 
obtained with $m_s \sim 150 $ MeV and 
$\gamma= \langle \bar{s} s \rangle / \langle \bar{u} u \rangle -1 \sim - 0.2$.
However, it was not always possible to calculate all 
physical quantities in QCD sum rules, especially those related to 
Goldstone bosons because of small momentum transfer and possible direct
instanton effects. 
However, by appropriately choosing the correlation
function and improving the continuum part, we can estimate effects of
explicit chiral symmetry breaking even for quantities related to the 
  Goldstone bosons. For example, in Ref.\cite{CCL96}
we calculated $g_{K N \Lambda}$, $g_{K N \Sigma}$ 
\footnote{A recent status on these couplings is given
in Ref.\cite{BMK97}.}
and compared to $g_{\pi NN}$,
and in Ref.\cite{CL97} we obtained
the decay constants $f_\pi$, $f_K$ and their ratio 
using the correlation function of the axial vector currents, for which 
no contamination from direct instantons is expected.

In this work, we proceed along these line by presenting a  
QCD sum rule calculation for the 
coupling constants 
 $g_{\pi \Lambda \Sigma}$ and $g_{K\Sigma \Xi}$ 
using the 3-point correlation function.  
Comparing these coupling constants  to each other
can provide further insight into 
SU(3) symmetry breaking effects on physical quantities
as in the case of $g_{KN\Lambda}$ and $g_{KN\Sigma}$.

In Sec. II and III we present 
sum rules for the coupling constants, taking into account the two
SU(3) symmetry breaking parameters,  $m_s$ and $\gamma$.
In Sec. IV we discuss uncertainties in our calculations and
the sign convention of the 
pole residues for ${1\over2}^+$ octet baryons,
and summarize our results.

\section{QCD sum rules for $g_{\pi \Lambda \Sigma}$}

We will closely follow the procedures
given in Refs.\cite{RRY83,RRY85,CCL96}.
Consider the three point function constructed of the two baryon
interpolating fields
 $\eta_B$, $\eta_{B'}$ and the pseudoscalar meson current $j_5$:
\begin{eqnarray}
\label{corr}
A(p,p',q) = \int dx\, dy\, \langle0| T (\eta_{B'}(x)j_5(y)
\overline{\eta}_B(0))|0
\rangle\, e^{i(p'\cdot x - q\cdot y)} .
\end{eqnarray}
In order to obtain $g_{\pi \Lambda \Sigma}$ we will use the following
interpolating
 fields for the $\Lambda$ and the $\Sigma$ as in Ref.\cite{RRY85,CCL96}:
\begin{eqnarray}
\eta_\Lambda&=&\sqrt{\frac{2}{3}} ~\epsilon_{abc} \left[
(u_a^T C\gamma_\mu s_b)\gamma_5\gamma^\mu d_c -
(d_a^T C\gamma_\mu s_b)\gamma_5\gamma^\mu u_c \right]  ,
\nonumber\\*
\eta_{\Sigma^\circ}
&=&\sqrt{2} ~~\epsilon_{abc} \left[
(u_a^T C\gamma_\mu s_b)\gamma_5\gamma^\mu d_c +
(d_a^T C\gamma_\mu s_b)\gamma_5\gamma^\mu u_c \right] ,
\end{eqnarray}
where u and d are the up and down quark fields, and $a,b,c$ are
color indices.
 $T$ denotes the transpose in Dirac space, and $C$ is the charge 
conjugation matrix.
 For the $\pi^0$ we choose the current
\begin{eqnarray}
j_{\pi^0} = \bar{u}i\gamma_5 u - \bar{d}i\gamma_5 d.
\end{eqnarray}
The sum rule after
Borel transformation in $p^2=p'^2$ is
\begin{eqnarray}
\lambda_\Lambda \lambda_\Sigma \frac{M_B}{M_\Sigma^2-M_\Lambda^2}
\left(e^{-M_\Lambda^2/M^2} - e^{-M_\Sigma^2/M^2}\right)
g_{\pi \Lambda \Sigma} \frac{f_\pi m_\pi^2}{\sqrt{2} m_q} =
\hspace{5cm}
\nonumber\\
\hspace{5cm}
- ~{2\over\sqrt{3}} \left(\frac{7}{12\pi^2} M^4
+ \frac{m_s^2}{4\pi^2} M^2
-~m_s \langle \bar{s} s \rangle \right)\langle \bar{q}q \rangle .
\label{eq:coup1}
\end{eqnarray}
Note that in this first exploratory work the pole-continuum transition
terms\cite{IS84,Ioffe95} have been neglected as was done in Ref.\cite{CCL96}.
 For $\lambda_\Lambda$ and $\lambda_\Sigma$, we use the
values obtained from the 
 following baryon sum rules for the $\Lambda$ and
 the $\Sigma$\cite{RRY85}:
\begin{eqnarray}
M^6  +\frac{2}{3}am_s(1-3\gamma) M^2+b M^2
+\frac{4}{9}a^2(3+4\gamma)=
2(2\pi)^4\lambda_\Lambda^2e^{-M_\Lambda^2/M^2} ,
\label{eq:lamlam}
\end{eqnarray}
\begin{eqnarray}
M^6-2am_s(1+\gamma)  M^2+b M^2
+\frac{4}{3}a^2=2(2\pi)^4 \lambda_\Sigma^2
e^{-M_\Sigma^2/M^2} ,
\label{eq:lamsig}
\end{eqnarray}
again paralleling the procedure of Ref.\cite{CCL96}.
Here $a \equiv -~(2\pi)^2 \langle\bar{q}q\rangle$,
 $b \equiv \pi^2 \langle (\alpha_s/ \pi) G^2 \rangle$,
and $\gamma \equiv \langle \bar{s}s \rangle / \langle 
\bar{q}q \rangle -1 \simeq -~0.2$.
We take the strange quark mass $m_s$ = 150 MeV,
and the pion decay constant $f_\pi$ = 133 MeV.
The sum rule in Eq.(\ref{eq:coup1}) 
does not display a plateau as a function of the Borel mass.
However, to
gain some idea of the SU(3) symmetry breaking effects we proceed by 
considering the value at Borel mass 
$M \simeq M_B=\frac{1}{2} (M_\Lambda + M_\Sigma)$, where
$M_\Lambda$ and $M_\Sigma$
are the masses of the $\Lambda$ and the $\Sigma$
particle respectively.
This approach parallels to that of Ref.\cite{CCL96} for 
 $g_{KN\Lambda}$ and $g_{KN\Sigma}$.
 At this Borel mass,  in the r.h.s. of Eq.(\ref{eq:coup1}), 
the contribution from the  s-quark mass correction is only 2\% of the 
 first term. 
Hence the leading order SU(3) breaking effects appear to be small.

Using the PCAC relation
$m_{\pi}^2 f_{\pi}^2 = - 4 m_q \langle\bar{q}q\rangle$,
we obtain
\begin{eqnarray}
g_{\pi \Lambda \Sigma} = 7.53
\end{eqnarray}
for $\langle\bar{q}q\rangle$ = -- (0.230 GeV)$^3$ and
$\langle (\alpha_s/ \pi) G^2 \rangle$ = (0.340 GeV)$^4$.
We now check the dependence of our result on
 the SU(3) symmetry breaking parameters.
If we take
$\langle\bar{s}s\rangle$ = 0.6 $\langle\bar{q}q\rangle$,
then the variation is within 0.3\%. In addition, for 
$m_s$ = 180 MeV the change of coupling constant is less than 0.8\%.
Thus, we find the coupling constant
$g_{\pi \Lambda \Sigma}$ is very weakly dependent on
the SU(3) symmetry breaking parameters.
One should be cautious, however, that larger SU(3) symmetry breaking
may be contained in higher order terms not considered here.

On the other hand, our result is more sensitive to different values of
the quark condensate.
We obtain $g_{\pi \Lambda \Sigma}$ = 7.35 and 7.13 for
$\langle\bar{q}q\rangle$ = -- (0.240 GeV)$^3$ and
-- (0.250 GeV)$^3$ respectively. 

\section{QCD sum rules for $g_{K\Sigma \Xi}$}

The interpolating fields of $\Sigma^+$ and $\Xi^\circ$ are defined
by\cite{RRY85}
\begin{eqnarray}
\eta_{\Sigma^+}&=&\epsilon_{abc}
(u_a^T C\gamma_\mu u_b)\gamma_5\gamma^\mu s_c,
\nonumber\\*
\eta_{\Xi^\circ}
&=& - \epsilon_{abc} (s_a^T C\gamma_\mu s_b)\gamma_5\gamma^\mu u_c ,
\end{eqnarray}
and we use
\begin{eqnarray}
j_{K^-} = \bar{s} i\gamma_5 u .
\end{eqnarray}
Then the final expression is
\begin{eqnarray}
\lambda_\Sigma \lambda_\Xi \frac{M_B}{M_\Xi^2-M_\Sigma^2}
\left(e^{-M_\Sigma^2/M^2} - e^{-M_\Xi^2/M^2}\right)
\sqrt{2}g_{K\Sigma \Xi} \frac{f_K m_K^2}{2 m_q} =
\hspace{5cm}
\nonumber\\
\hspace{5cm}
+ \left(\frac{9}{10\pi^2} M^4
+ \frac{7m_s^2}{5\pi^2} M^2
- {6\over 5} ~m_s \langle\bar{s}s\rangle \right) \langle\bar{q}q\rangle .
\label{eq:coup2}
\end{eqnarray}
In this case the contribution of the s-quark mass corrections
is also small; about 3\% of the first term.  

For $\lambda_\Xi$, we use the following sum rule for $\Xi$\cite{RRY85}:
\begin{eqnarray}
M^6+b M^2+\frac{4}{3}a^2(1+\gamma)^2=2(2\pi)^4 \lambda_\Xi^2
e^{-M_\Xi^2/M^2} .
\end{eqnarray}
Then, the value of $g_{K\Sigma \Xi}$ is
\begin{eqnarray}
g_{K\Sigma \Xi} = -~7.02
\end{eqnarray}
for $\langle\bar{q}q\rangle$ = -- (0.230 GeV)$^3$ and $f_k$ = 160 MeV.
The variation of the coupling constant is within 0.5\% if we
take $\langle\bar{s}s\rangle$ =  0.6 $\langle\bar{q}q\rangle$.
On the other hand,
we obtain $g_{K\Sigma \Xi}$ = -- 7.87 and -- 8.75 for
$\langle\bar{q}q\rangle$ = -- (0.240 GeV)$^3$ and
-- (0.250 GeV)$^3$ respectively.  
In addition the coupling constant is rather dependent on 
the s-quark mass. For example, if we take $m_s$ = 180 MeV, then 
$g_{K\Sigma \Xi}$ = --~8.55 for 
$\langle\bar{q}q\rangle$ = -- (0.230 GeV)$^3$. 
In the case of $g_{K\Sigma \Xi}$ we insert the values of the s-quark mass 
in the l.h.s. of Eq.(\ref{eq:coup2}) and the quark condensate 
in the r.h.s. directly 
instead of using the PCAC relation for the kaon.
Therefore the variation of the coupling constant is much larger than that
for the case of $g_{\pi \Lambda \Sigma}$. 

\section{Discussion}

In the previous sections we calculated
$g_{\pi \Lambda \Sigma}$ and $g_{K\Sigma \Xi}$ by following the same
procedures in Ref.\cite{CCL96}.
Here we discuss contributions not included
in the previous calculations;
First, the next-leading operator is dimension five 
$\langle g_s \bar{q} \sigma \cdot G q \rangle$, and 
it may contribute to the OPE side with considerable weight
as in nucleon mass sum rules\cite{Leinweber97}.
In addition, operators of dimension seven 
may also be important in the OPE side
as a further power correction.
Second, the pole-continuum transition terms 
are neglected as we said previously
\footnote{In the case of $g_{\pi NN}$
its contribution is at most 5\%\cite{BK96-1,BK96-2}.}.
Last, the contribution of pure continuum is not included.

While the inclusion of higher order power corrections would 
significantly complicate the exploratory analysis presented here,
one can easily include the pure continuum contribution by 
considering
the following factor in the OPE side:
\begin{eqnarray}
E_i = 1 - \sum_{k=0}^i \frac{s_0^k}{k~! ~(M^2)^k} ~e^{-\frac{s_0}{M^2}} ,
\end{eqnarray}
where $s_0$ is a continuum threshold.
For example, including the effect of the pure continuum
Eq.(\ref{eq:coup1}) becomes
\begin{eqnarray}
\lambda_\Lambda \lambda_\Sigma \frac{M_B}{M_\Sigma^2-M_\Lambda^2}
\left(e^{-M_\Lambda^2/M^2} - e^{-M_\Sigma^2/M^2}\right)
g_{\pi \Lambda \Sigma} \frac{f_\pi m_\pi^2}{\sqrt{2} m_q} =
\hspace{5cm}
\nonumber\\
\hspace{5cm}
- ~{2\over\sqrt{3}} \left(\frac{7}{12\pi^2} E_1 M^4
+ \frac{m_s^2}{4\pi^2} E_0 M^2
-~m_s \langle \bar{s} s \rangle \right)\langle \bar{q}q \rangle ,
\label{eq:coup1_2}
\end{eqnarray}
and Eqs.(\ref{eq:lamlam}), (\ref{eq:lamsig}) can be written as
\begin{eqnarray}
E_2 M^6  +\frac{2}{3}am_s(1-3\gamma) E_0 M^2+b E_0 M^2
+\frac{4}{9}a^2(3+4\gamma)=
2(2\pi)^4\lambda_\Lambda^2e^{-M_\Lambda^2/M^2} ,
\label{eq:lamlam_2}
\end{eqnarray}
\begin{eqnarray}
E_2 M^6-2am_s(1+\gamma) E_0 M^2+b E_0 M^2
+\frac{4}{3}a^2=2(2\pi)^4 \lambda_\Sigma^2
e^{-M_\Sigma^2/M^2} ,
\label{eq:lamsig_2}
\end{eqnarray}
where we assume the same continuum threshold
in Eqs.(\ref{eq:coup1_2}), (\ref{eq:lamlam_2}) and (\ref{eq:lamsig_2}).
In Table I we present our previous results of
 $g_{KN\Lambda}$, $g_{KN\Sigma}$ 
and the present results of $g_{\pi \Lambda \Sigma}$, $g_{K\Sigma \Xi}$ 
without (and with) a continuum model. The previous analysis of 
Ref.\cite{CCL96}
has been repeated with the continuum model correction.
We take the continuum threshold to be $s_0$=2.560 GeV$^2$,
2.756 GeV$^2$, and 2.856 GeV$^2$ 
for the case of  
 $g_{KN\Lambda}$, $g_{KN\Sigma}$ (and $g_{\pi \Lambda \Sigma}$),
and $g_{K\Sigma \Xi}$ respectively
considering the next $\Lambda$(1600),
$\Sigma$(1660), and $\Xi$(1690) particle each other, although
in the case of $\Xi$(1690) its quantum number is not clarified
in experiments\cite{PRD96}.
In our calculation the continuum contribution is always less than 50\%
of the phenomenological side at the relevant Borel mass.
A comparison to
fitting analyses of experimental data\cite{TRS95,Na79Du83}
is also provided.

In the table the first row is a prediction from SU(3) relations between  
meson-baryon coupling constants. 
The SU(3) symmetry, using de Swart's convention\cite{Swart63},
predicts
\begin{eqnarray}
g_{KN\Lambda} &=& - ~\frac{1}{\sqrt{3}} (3 - 2\alpha_D) g_{\pi NN} ,
\nonumber \\*
g_{KN\Sigma} &=& + ~(2\alpha_D -1) g_{\pi NN} ,
\nonumber\\*
g_{\pi \Lambda \Sigma} &=& {2\over\sqrt{3}} \alpha_D g_{\pi NN} ,
\nonumber\\*
g_{K\Sigma \Xi} &=& - g_{\pi NN} ,
\label{eq:su3}
\end{eqnarray}
where  $\alpha_D$ is the fraction of the D type coupling,
 $\alpha_D = \frac{D}{D+F}$.
We take $\alpha_D$ from a recent analysis of hyperon semi-leptonic
decay data by Ratcliffe, $\alpha_D$=0.64\cite{Ratcliffe9697}
while 7/12 in the SU(3) symmetric limit\cite{RRY83},
and $g_{\pi NN}$ from an analysis of the $np$ data 
by Ericson {\it et al.}\cite{ELBO96}, $g_{\pi NN}$=13.43.
We denote the error bar allowing for SU(3) symmetry breaking 
at the 20\% level.

One can see that the results with a continuum
model are larger than those without the continuum contribution.   
However, the corrections range from 20 to 45\% and suggest that further
analysis is required before any firm conculsions may be drawn.
Full quantitative analysis along the lines of 
Leinweber's work\cite{Leinweber97} 
would require all of the above mentioned corrections and is beyond 
the scope of this first exploratory calculation.

Let us comment on the sign convention of $\lambda_B$.
Usually we construct the  interpolating fields for the octet baryons 
by starting from the nucleon current and then making SU(3) rotations.
Then the phase will be the same for all baryon states 
assuming exact SU(3) symmetry.
But, in the real world the $\lambda_B$s are not SU(3) symmetric
and the phase can be changed according to the level of
SU(3) symmetry breaking.
However, our previous calculation of
 $g_{KN\Lambda}$ and $g_{KN\Sigma}$, and the present
calculation of  $g_{\pi \Lambda \Sigma}$ and $g_{K\Sigma \Xi}$
show that contribution of the s-quark mass corrections is very small
compared to the leading term, and thus the relative signs of $\lambda_B$
are the same for all octet baryons.

One can easily check this as below.
In Ref.\cite{CCL96} the coupling constants in our diagram correspond
to -- $g_{KN\Lambda}$ and -- $g_{KN\Sigma}$ respectively according to
de Swart's sign convention\cite{Swart63}.
Then, our results can be rewritten as follows:
\begin{eqnarray}
 g_{KN\Lambda} &\simeq& {- \over \lambda_N \lambda_\Lambda} ,
\nonumber\\*
 g_{KN\Sigma}  &\simeq& {+ \over \lambda_N \lambda_\Sigma} ,
\label{eq:sign-1}
\end{eqnarray}
where + and -- in the r.h.s. mean that the signs of numerators
are + and --  respectively.
Similarly our present results for
$g_{\pi \Lambda \Sigma}$ and $g_{K\Sigma \Xi}$ give
\begin{eqnarray}
 g_{\pi \Lambda \Sigma} &\simeq& {+ \over \lambda_\Lambda \lambda_\Sigma} ,
\nonumber\\*
 g_{K\Sigma \Xi}  &\simeq& {- \over \lambda_\Sigma \lambda_\Xi} .
\label{eq:sign-2}
\end{eqnarray}
Assuming $g_{\pi NN} > 0$, 
one can see that the relative signs of $\lambda_B$s are the same
as can be seen by comparing Eqs.(\ref{eq:sign-1}) and (\ref{eq:sign-2}) to 
 Eq.(\ref{eq:su3}).
This result follows from the fact that
the SU(3) symmetry is slightly broken in our sum rules.
In fact, there is another sign convention for meson-baryon
coupling constants\cite{Na79Du83}.
As emphasized in Ref.\cite{AS90}, however, both conventions lead to the same
result for the only physically meaningful sign, $g_{KN\Lambda}$
and $g_{KN\Sigma} \cdot \mu(\Sigma^\circ \Lambda)$.
Where, $\mu(\Sigma^\circ \Lambda)$ is the $\Sigma^\circ - \Lambda$
transition moment.

In summary, using the 3-point correlation function method
$g_{\pi \Lambda \Sigma}$ and $g_{K\Sigma \Xi}$
are obtained in the QCD sum rule approach.
In both cases the contribution of SU(3) breaking effects 
in the leading order OPE side is less than 5\%.
The pattern of SU(3) breaking
appears to be different from that based on SU(3) relations.
Omission of continuum model contributions,
as done in previous calculations, appears to be too crude.
The couplings increase when the continuum model corrections are
included, in some cases by nearly 50\%.
It would be interesting to further refine the QCD sum rule approach 
to allow a more depth study.

\acknowledgements

The author thanks Prof. Su H. Lee for valuable discussions.
He is also grateful to Prof. A.W. Thomas, Dr. A.G. Williams, 
Dr. D.B. Leinweber for 
useful discussions and comments, and especially to Dr. D.B. Leinweber
for his comments on the manuscript.
The author wishes to acknowledge the financial support of 
Korea Research Foundation (KRF) made in the program year 1997.
This work is supported in part by KOSEF
through CTP at Seoul National University and in part by
Special Research Centre for the Subatomic Structure of Matter
in University of Adelaide.

\newpage
\begin{table}

Table I. Coupling constants. 

\vspace{0.5cm}
\begin{tabular}{c c c c c}
\\
Coupling Constants & $g_{KN\Lambda}$          & $g_{KN\Sigma}$ 
                   & $g_{\pi \Lambda \Sigma}$ & $g_{K\Sigma \Xi}$ \\ \\
\hline
\\
SU(3)              & -- 16.01 $\sim$ -- 10.67 & 3.01 $\sim$ 4.51 
                   & 7.94 $\sim$ 11.90        & -- 16.12 $\sim$ -- 10.74  \\ \\
QSR (w/o Cont.)$^\dagger$    & -- 6.96                  & 1.05
                   & 7.53                     & -- 7.02 \\ \\
QSR (with  Cont.)$^\dagger$    & -- 8.34                  & 1.26
                   & 10.79                    & -- 10.22 \\ \\
Exp. Fit           & -- 13.68$^1$             & ~3.86$^1$
                   & ~11.75$^2$               & ~~N/A        \\ \\
\end{tabular}
\vspace{0.5cm}
$^\dagger$ We take $\langle\bar{q}q\rangle$ = -- (0.230 GeV)$^3$.\\
$^1$ Ref.\cite{TRS95}. \\
$^2$ Ref.\cite{Na79Du83}. \\ 
\end{table}
 


\end{document}